\documentclass[letter,aps,pra,amsmath,amssymb,twocolumn,nofootinbib,superscriptaddress]{revtex4-1}
\usepackage[utf8]{inputenc}
\usepackage[T1]{fontenc}
\usepackage{graphicx}
\usepackage{dcolumn}
\graphicspath{{images/}}
\usepackage[colorlinks,linkcolor=blue,citecolor=blue,urlcolor=blue]{hyperref}

\begin{document}

\begin{abstract}
  Entangling a mechanical oscillator with an optical mode is an enticing and yet a very challenging goal in cavity-optomechanics. Here we consider a pulsed scheme to create EPR-type entanglement between a travelling-wave light pulse and a mechanical oscillator. The entanglement can be verified unambiguously by a pump--probe sequence of pulses. In contrast to schemes that work in a steady-state regime under continuous wave drive this protocol is not subject to stability requirements that normally limit the strength of achievable entanglement. We investigate the protocol's performance under realistic conditions, including mechanical decoherence, in full detail. We discuss the relevance of a high mechanical $Q\cdot f$-product for entanglement creation and provide a quantitative statement on which magnitude of the $Q\cdot f$-product is necessary for a successful realization of the scheme. We determine the optimal parameter regime for its operation, and show it to work in current state-of-the-art systems.
\end{abstract}
\title{Quantum entanglement and teleportation in pulsed cavity-optomechanics}
\author{Sebastian G.\ Hofer}
\email{sebastian.hofer@univie.ac.at}
\affiliation{Vienna Center for Quantum Science and Technology (VCQ), Faculty of Physics, University of Vienna, Boltzmanngasse 5, 1090 Vienna, Austria}
\affiliation{Institute for Theoretical Physics, Institute for Gravitational Physics, Leibniz University Hannover, Callinstra\ss{}e 38, 30167 Hannover, Germany}
\author{Witlef Wieczorek}
\affiliation{Vienna Center for Quantum Science and Technology (VCQ), Faculty of Physics, University of Vienna, Boltzmanngasse 5, 1090 Vienna, Austria}
\author{Markus Aspelmeyer}
\affiliation{Vienna Center for Quantum Science and Technology (VCQ), Faculty of Physics, University of Vienna, Boltzmanngasse 5, 1090 Vienna, Austria}
\author{Klemens Hammerer}
\affiliation{Institute for Theoretical Physics, Institute for Gravitational Physics, Leibniz University Hannover, Callinstra\ss{}e 38, 30167 Hannover, Germany}
\keywords{optomechanics}
\pacs{}
\maketitle

\section{Introduction}
\label{sec-1}

In optomechanical systems a cavity mode can be strongly coupled to a high-quality mechanical oscillator via radiation pressure or dipole gradient forces \cite{marquardt_optomechanics_2009,kippenberg_cavity_2008,regal_cavity_2011}. Quantum effects \cite{aspelmeyer_quantum_2010-1,genes_chapter_2009,verlot_optomechanical_2007} are starting to play an increasingly important role: In the microwave regime ground-state cooling via laser-cooling techniques \cite{teufel_sideband_2011}, strong coupling \cite{teufel_circuit_2011,oconnell_quantum_2010} and coherent control of single-phonon excitations \cite{oconnell_quantum_2010} have been successfully achieved. In the optical regime cooling to the quantum ground state \cite{chan_laser_2011} and effects of strong coupling \cite{groeblacher_observation_2009} have been demonstrated in recent experiments. It is as yet an outstanding goal to observe genuine quantum effects such as entanglement \cite{schroedinger_gegenwaertige_1935} at macroscopic length and mass scales.

Entanglement of a mechanical oscillator with light has been predicted in a number of theoretical studies \cite{paternostro_creating_2007,genes_robust_2008,miao_universal_2010,vitali_optomechanical_2007,galve_bringing_2010,genes_simultaneous_2008,vitali_entangling_2007,ghobadi_optomechanical_2011,ghobadi_quantum_2011,abdi_effect_2011,mari_gently_2009} and would be an intriguing demonstration of optomechanics in the quantum regime. These studies, as well as similar ones investigating entanglement among several mechanical oscillators \cite{mancini_entangling_2002,zhang_quantum-state_2003,pinard_entangling_2005,pirandola_macroscopic_2006,hartmann_steady_2008,vacanti_optomechanical_2008,huang_entangling_2009,vitali_stationary_2007,ludwig_entanglement_2010}, explore entanglement in the \emph{steady-state regime}. In this regime the optomechanical system is driven by one or more continuous-wave light fields and settles into a stationary state, for which the interplay of optomechanical coupling, cavity decay, damping of the mechanical oscillator, and thermal noise forces may remarkably give rise to persistent entanglement between the intracavity field and the mechanical oscillator.

Entanglement in the steady-state regime shows two main characteristic features: Firstly, entanglement reaches a maximal value when the system is driven close to a point of dynamical instability. \textit{Vice versa}, the limits on the strength of entanglement achievable in protocols working in the steady-state regime are set by the very conditions guaranteeing a dynamically stable, stationary state. Recent studies indicate that these limitations can even become rather restrictive when a finite laser linewidth is taken into account \cite{ghobadi_optomechanical_2011,abdi_effect_2011}. Secondly, the verification of entanglement between the intracavity field and the moving mirror has to be performed via measurements on the outcoupled light leaving the optomechanical system. Ultimately only correlations between modes of the light field are measured, from which any entanglement involving the mechanical oscillator has to be inferred. However, due to the curious feature of quantum correlations that ``no entanglement is necessary to distribute entanglement'' \cite{cubitt_separable_2003,mita_distribution_2008}, this sort of inference is in general a delicate issue. It is unambiguously only possible under additional assumptions regarding the particular dynamics (\ie{}, the system's Hamiltonian) and structure of the steady state \cite{paternostro_creating_2007}.

An alternative approach to achieving optomechanical entanglement is to work in the \emph{pulsed regime}, where entanglement is created and verified with two subsequent pulses of light. This strategy has first been developed in the context of atomic ensembles \cite{hammerer_teleportation_2005} and was recently considered for systems employing levitated microspheres trapped in an optical cavity \cite{romero-isart_optically_2011}. A pulsed scheme does not rely on the existence of a stable steady state, which provides us with the benefit that entanglement is not limited by stability requirements. The temporal ordering of the pulses excludes the possibility of distributing entanglement without using entanglement \cite{cubitt_separable_2003,mita_distribution_2008} such that it provides a direct (\ie{}, without additional assumptions) and unambiguous test of entanglement. Similar protocols have also been discussed for micromirrors in free space (\ie{}, without the use of an optical cavity) \cite{mancini_scheme_2003,pirandola_continuous-variable_2003}.

In this article we provide a complete treatment of a protocol for the generation and verification of optomechanical entanglement using pulsed light. In addition to an idealized scenario, which was briefly discussed in \cite{romero-isart_optically_2011}, we include in our description the full dynamics of the optomechanical system. Our derivation provides an exhaustive discussion of imperfections, how they affect the performance of the protocol and how these effects can be minimised. Most prominently, we (perturbatively) include thermal decoherence of the mechanical system and find that a high $Q \cdot f$-product (quality factor times the frequency of the mechanical oscillator) plays a crucial role in the creation of optomechanical entanglement. More specifically, we find that the relation $Q\cdot f\gg k_{\mathrm{B}}T/h$ (where $T$ is the temperature of the mechanical environment) has to hold, which is a general and often very stringent condition to observe quantum effects in optomechanical systems \cite{chang_ultrahigh-q_2010}. A large $Q\cdot f$-product is thus one of the most important aspects to consider in the design of novel high-quality mechanical resonators.

To further explore the effect of imperfections on the protocol, we optimise the amount of created entanglement with respect to key experimental parameters and present specific values for two existing optomechanical systems. We find that creation of entanglement is possible in a parameter regime which is realistic yet challenging for current state-of-the-art setups.
Very importantly, our treatment also provides a \emph{quantitative} statement on what magnitude of the $Q\cdot f$-product is necessary in order to successfully create entanglement for a given bath temperature $T$. These findings provide a general understanding of the requirements to observe quantum effects in optomechanical systems and represent essential information for the material development and for the further design of future optomechanical structures.

We finally note that the quantum state created in this protocol exhibits a type of entanglement known as Einstein--Podolsky--Rosen (EPR) entanglement \cite{einstein_can_1935} between the mechanical oscillator and the light pulse \cite{romero-isart_optically_2011}. It thus provides the canonical resource for quantum information protocols involving continuous variable (CV) systems \cite{braunstein_quantum_2005,braunstein_quantum_2003}. We give a detailed description of how optomechanical EPR entanglement can be used for the teleportation of the state of a propagating light pulse onto a mechanical oscillator.

Note that other promising perspectives for \emph{pulsed optomechanics} have recently been discussed also in \cite{vanner_pulsed_2011,wang_ultraefficient_2011,cerrillo_pulsed_2011,vanner_selective_2011,romero-isart_large_2011}, albeit in a very different parameter regime employing light pulses which are short on the time scale of a mechanical oscillation. The importance of temporal ordering in the verification of optomechanical entanglement has also been pointed out in~\cite{miao_universal_2010}.

The paper is organized as follows: \Sref{sec:motivation} contains the main results of this work. We first describe entanglement creation under idealized circumstances and outline a way to verify it unambiguously. Additionally, we show how it can be used as a resource for CV teleportation. In \sref{sec:optimised} we analyze the influence of imperfections on the protocol's performance and find the optimal parameter regime for maximal entanglement. \Sref{sec:completemodel} gives a detailed description of the full system dynamics. The Appendix contains a short derivation of the effective system Hamiltonian in the pulsed regime.
\section{Central results}
\label{sec-2}
\label{sec:results}
\subsection{Motivation for the pulsed scheme}
\label{sec-2-1}
\label{sec:motivation}
\subsubsection{Cavity optomechanical system}
\label{sec-2-1-1}
\label{sec:system}

\label{sec-2-1-1-1}

\begin{figure}
  \includegraphics{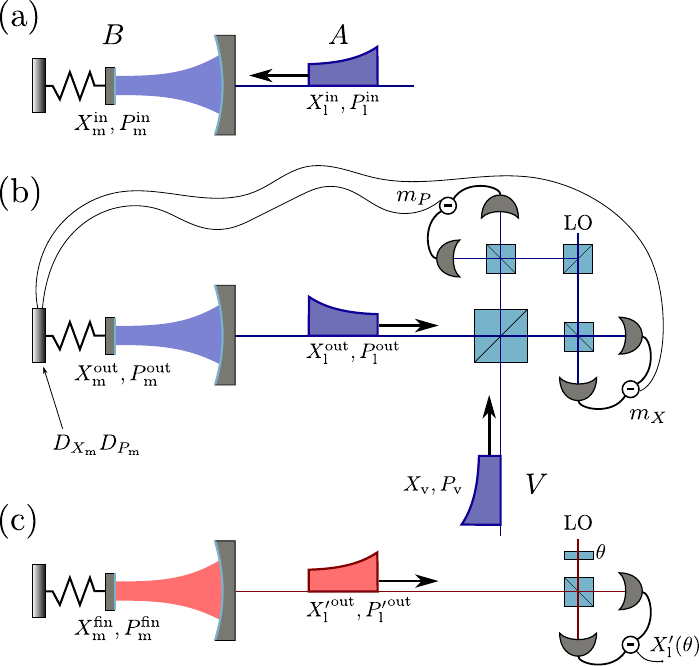}
  \caption{(Color online) Schematic of the system and the teleportation protocol: (a) A blue detuned light pulse (A) is entangled with the mirror (B). (b) A second light pulse (V) is prepared in the input state and interferes with A on a beam splitter. Two homodyne detectors measure $P^{\mathrm{out}}_{\mathrm{l}}+X_{\mathrm{v}}$ and $X^{\mathrm{out}}_{\mathrm{l}}+P_{\mathrm{v}}$, yielding outcomes $m_{X}$ and $m_{P}$, respectively. Feedback is applied by displacing the mirror state in phase space by a unitary transformation $D_{X_{\mathrm{m}}}\!(m_X)\,D_{P_{\mathrm{m}}}\!(m_P)$. (c) To verify the success of the protocol, the mirror state is coherently transferred to a red detuned laser pulse and a generalised quadrature $X'_{\mathrm{l}}(\theta) ={X'_{\mathrm{l}}}^{\mathrm{out}} \cos \theta + {P_{\mathrm{l}}'}^{\mathrm{out}} \sin \theta$ is measured. Repeating steps (a)--(c) for the same input state but for different phases $\theta$ yields a reconstruction of the mirror's quantum state.}
  \label{fig:protocol}
\end{figure}
Let us consider an optomechanical cavity in a Fabry--Pérot-type setup (\fref{fig:protocol}), with mechanical oscillation frequency $\om$, mechanical dissipation rate $\gamma$, optical resonance frequency $\omega_{\mathrm{c}}$ and cavity decay rate $\kappa$. A light pulse of duration $\tpulse$ and carrier frequency $\omega_{\mathrm{l}}$ impinges on the cavity and interacts with the oscillatory mirror mode via radiation pressure.

\label{sec-2-1-1-2}

In a frame rotating with the laser frequency, the system is described by the (effective) Hamiltonian \cite{mancini_quantum_1994}
\begin{multline}
  \label{eq:1}
  H=\om^{} \amd\am + \detuning^{} \acd\ac + g \left( \am + \amd \right)\left( \ac + \acd \right)
\end{multline}
where $\am$ and $\ac$ are annihilation operators of the mechanical and optical mode, respectively. The conditions under which \eqref{eq:1} is valid are discussed in detail in \sref{sec:linearization}. The first two terms give the energy of the mechanical oscillator and the cavity field, where $\detuning=\omega_{\mathrm{c}}-\omega_{\mathrm{l}}$ is the detuning of the laser drive with respect to the cavity resonance. The last term describes the linearised optomechanical coupling (with coupling constant $g$) via radiation pressure $x_{\mathrm{m}}x_{\mathrm{c}} \propto \left( \am + \amd \right)\left( \ac + \acd \right) = \left( \am \acd + \amd \ac \right) + \left( \am \ac + \amd \acd \right)$, which can be decomposed into two terms \cite{aspelmeyer_quantum_2010-1,mancini_scheme_2003,pirandola_continuous-variable_2003,zhang_quantum-state_2003}: a beam-splitter-like interaction (the first term) and a two-mode-squeezing interaction (the second term). The former can be used to cool the mirror as well as to generate a state swap between the mechanical and the optical mode, while the latter term describes the optomechanical analogue to the optical down-conversion process in an optical parametric amplifier and is known to create entanglement from coherent input states \cite{ou_realization_1992,aspelmeyer_quantum_2010-1}.

We assume the pulse to approximately be a flat-top pulse, which has a constant amplitude for the largest part, but possesses a smooth head and tail (see the Appendix). The coupling constant $g$ is then given by
\begin{equation}
  \label{eq:2}
  g=g_0 \sqrt{\frac{2\kappa}{\detuning^2+\kappa^2}\frac{N_{\mathrm{ph}}}{\tpulse}},
\end{equation}
with $N_{\mathrm{ph}}$ being the number of photons in the pulse (see \sref{sec:linearization}). The single-photon coupling constant $g_0$ is defined by $g_0= \omega_{\mathrm{c}} x_0/L$, where $x_{0}$ is the size of the zero-point-motion of the mechanical oscillator and $L$ is the cavity length.
It is possible to make a single one of the interaction terms dominant by tuning the laser such, that one of its motional sidebands $\omega_{\mathrm{l}}\pm \om$ is resonant with the cavity, where for blue detuning the resonant scattering to the lower (Stokes) sideband ($\omega_{\mathrm{c}}=\omega_{\mathrm{l}} - \om$) enhances the down-conversion interaction, while for red detuning the resonant scattering to the upper (anti-Stokes) sideband ($\omega_{\mathrm{c}}=\omega_{\mathrm{l}} + \om$) enhances the beam-splitter interaction \cite{aspelmeyer_quantum_2010-1}.
In this pump-probe scheme we make use of both dynamics separately: Pulses tuned to the blue side of the cavity resonance are applied to create entanglement, while pulses on the red side are later used to read out the final mirror state. A similar separation of Stokes and anti-Stokes sideband was suggested in \cite{mancini_scheme_2003,pirandola_continuous-variable_2003} by selecting different angles of reflection of a light pulse scattered from a vibrating mirror in free space.

\label{sec-2-1-1-3}

The full system dynamics, including the dissipative coupling of the mirror and the cavity decay, are described by quantum Langevin equations \cite{gardiner_quantum_2004}, which determine the time evolution of the corresponding operators $x_{\mathrm{m}}=(\am+\amd)/\sqrt{2}$, $p_{\mathrm{m}}=-\rmi(\am-\amd)/\sqrt{2}$ and $\ac,\acd$. They read
\begin{subequations}
  \label{eq:3}
  \begin{align}
    \label{eq:3a}
    \dot{x}_\mathrm{m}&=\phantom{-}\om p_{\mathrm{m}},\\
    \label{eq:3b}
    \dot{p}_\mathrm{m}&=-\om x_{\mathrm{m}} - \gamma\, p_{\mathrm{m}}- \sqrt{2}\, g \left( \ac + \acd \right) - \sqrt{2\gamma}\, f,\\
    \label{eq:3c}
    \dot{a}_\mathrm{c}&=-(\rmi\detuning^{}+\kappa) \ac - \rmi \sqrt{2}\, g\, x_{\mathrm{m}} - \sqrt{2\kappa}\, \ain,
  \end{align}
\end{subequations}
where we introduced the (self-adjoint) Brownian stochastic force $f$, and quantum noise $\ain$ entering the cavity from the electromagnetic environment. Both $\ain$ and---in the high-temperature limit---$f$ are assumed to be Markovian. Their correlation functions are thus given by $\langle{\ain(t) \aind(t')}\rangle=\delta(t-t')$ (in the optical vacuum state) and $\mean{f(t)f(t')+f(t')f(t)}=(2\bar{n}+1)\delta(t-t')$ (in a thermal state of the mechanics) \cite{gardiner_quantum_2004}.
\subsubsection{Creation of optomechanical entanglement}
\label{sec-2-1-2}
\label{sec:entanglement}

\label{sec-2-1-2-1}

In this section we impose the following conditions on the system's parameters. Firstly, we drive the cavity with a blue detuned laser pulse ($\detuning=-\om$) and assume to work in the resolved-sideband regime ($\kappa\ll\om$) to enhance the down-conversion dynamics. Note that in this regime a stable steady state only exists for very weak optomechanical coupling \cite{ludwig_optomechanical_2008}, which poses a fundamental limit to the amount of entanglement that can be created in a continuous-wave scheme \cite{genes_robust_2008}. In contrast, a pulsed scheme does not suffer from these instability issues. In fact, it is easy to check by integrating the full dynamics (see \sref{sec:fullequations}) up to time $\tpulse$,
that working in this particular regime yields maximal entanglement, which increases with increasing sideband resolution $\om/\kappa$. Secondly we assume a weak optomechanical coupling $g\ll \kappa$, such that only first-order interactions of photons with the mechanics contribute. This minimises pulse distortion and simplifies the experimental realization of the protocol. Taken together, the conditions $g\ll\kappa\ll\om$ allow us to invoke the rotating-wave approximation (RWA), which amounts to neglecting the beam-splitter term in \eqref{eq:1}. Also, we neglect mechanical decoherence effects in this section. We emphasise that this approximation is justified as long as the total duration of the protocol is short compared to the effective mechanical decoherence time $1/\gamma \bar{n}$, where $\gamma$ is the mechanical damping rate and $\bar{n}$ is the thermal occupation of the corresponding bath. Corrections to this simplified model---including the treatment of mechanical decoherence and dynamics beyond the RWA---will be addressed in \sref{sec:optimised}.

\label{sec-2-1-2-2}

Based on the assumptions above we can now simplify equations \eqref{eq:3}. For convenience we go into a frame rotating with $\om$ by substituting $\ac\rightarrow \ac\e^{\rmi \om t}$, $\ain \rightarrow \ain \e^{\rmi \om t}$ and $\am\rightarrow \am\e^{-\rmi \om t}$. Note that in this picture the central frequency of $\ain$ is located at $\omega_{\mathrm{l}}-\om=\omega_{\mathrm{c}}$. In the RWA the Langevin equations then simplify to
\begin{subequations}
  \label{eq:4}
  \begin{align}
    \label{eq:5a}
    \dot{a}_\mathrm{c}&=-\kappa \ac\, - \rmi g\, \amd - \sqrt{2\kappa}\, \ain,\\
    \label{eq:5b}
    \dot{a}_\mathrm{m}&=- \rmi g\, \acd.
  \end{align}
\end{subequations}

\label{sec-2-1-2-3}

In the limit $g\ll \kappa$ we can use an adiabatic solution for the cavity mode and we therefore find
\begin{subequations}
  \label{eq:5}
  \begin{align}
    \label{eq:6a}
    \ac(t) &\approx -\rmi \frac{g}{\kappa} \amd(t) - \sqrt{\frac{2}{\kappa}} \ain(t),\\
    \label{eq:6}
    \am(t) &\approx \e^{G t} \am(0) +\rmi \sqrt{2G} \e^{Gt} \intdb{0}{t}{s} \e^{-Gs} \aind(s),
  \end{align}
\end{subequations}
where we defined $G=g^2/\kappa$. Equation \eqref{eq:6} shows that the mirror motion gets correlated to a light mode of central frequency $\omega_\mathrm{l}-\om$ (which coincides with the cavity resonance frequency $\omega_{\mathrm{c}}$) with an exponentially shaped envelope $\alpha_{\mathrm{in}}(t) \propto \e^{-Gt}$. Using the standard cavity input-output relations $\aout=\ain+\sqrt{2\kappa}\,\ac$ allows us to define a set of normalised temporal light modes
\begin{subequations}
  \label{eq:7}
  \begin{align}
    \label{eq:7a}
    A_{\mathrm{in}}&=\sqrt{\frac{2G}{1-\e^{-2G \tpulse}}}\intdb{0}{\tpulse}{t} \e^{-G t}\ain(t),\\
    \label{eq:7b}
    A_{\mathrm{out}}&=\sqrt{\frac{2G}{\e^{2G\tpulse}-1}}\intdb{0}{\tpulse}{t} \e^{G t}\aout(t),
  \end{align}
\end{subequations}
which obey the canonical commutation relations $[A_{i},A^{\dagger}_{i}]=1$.
Together with the definitions $B_{\mathrm{in}}=\am(0)$ and $B_{\mathrm{out}}=\am(\tpulse)$ we arrive at the following expressions, which relate the mechanical and optical mode at the end of the pulse $t=\tpulse$
\begin{subequations}
  \label{eq:8}
  \begin{align}
    \label{eq:8a}
    A_{\mathrm{out}}&=-\e^{G\tpulse} A_{\mathrm{in}} -\rmi \sqrt{\e^{2G\tpulse}-1}B_{\mathrm{in}}^{\dagger},\\
    \label{eq:8b}
    B_{\mathrm{out}}&=\e^{G\tpulse} B_{\mathrm{in}} + \rmi \sqrt{\e^{2G\tpulse}-1}A_{\mathrm{in}}^{\dagger}.
  \end{align}
\end{subequations}
By expressing equations \eqref{eq:8} in terms of quadratures $X_\mathrm{m}^i=(B_i+B_i^\dagger)/\sqrt{2}$ and $X_\mathrm{l}^i=(A_i+A_i^\dagger)/\sqrt{2}$, where $i \in\{\mathrm{in},\mathrm{out}\}$, and their corresponding conjugate variables, we can calculate the so-called EPR variance $\depr$ of the state after the interaction.
For light initially in vacuum $(\Delta X_\mathrm{l}^\mathrm{in})^2=(\Delta P_\mathrm{l}^\mathrm{in})^2=\frac{1}{2}$ and the mirror in a thermal state $(\Delta X_\mathrm{m}^\mathrm{in})^2=(\Delta P_\mathrm{m}^\mathrm{in})^2=n_0+\frac{1}{2}$, the state is entangled iff \cite{duan_inseparability_2000}
\begin{equation}
  \begin{split}
    \label{eq:11}
    \depr&=\left[\Delta(X^{\mathrm{out}}_{\mathrm{m}}+P^{\mathrm{out}}_{\mathrm{l}})\right]^2+\left[\Delta(P^{\mathrm{out}}_{\mathrm{m}}+X^{\mathrm{out}}_{\mathrm{l}})\right]^2\\
    &=2(n_0+1)\left(\e^r-\sqrt{\e^{2r}-1}\right)^2<2,
  \end{split}
\end{equation}
where  $r=G\tpulse$ is the squeezing parameter and $n_0$ the initial occupation number of the mechanical oscillator.
Note that in the limit of large squeezing $r\gg 1$ we find that the variance $\depr\approx (n_{0}+1)\e^{-2r}/2$ is suppressed exponentially, which shows that the created state asymptotically approximates an EPR state. Therefore, this state can be readily used to conduct optomechanical teleportation as described in \sref{sec:protocol}.

\label{sec-2-1-2-4}

Rearranging \eqref{eq:11}, we find that the state is entangled as long as
\begin{equation}
  \label{eq:12}
  r>r_0=\frac{1}{2}\ln\left(\frac{(n_0+2)^2}{4(n_0+1)}\right) \overset{n_0\rightarrow \infty}{\sim} \frac{1}{2}\ln{n_0}.
\end{equation}
This illustrates that in our scheme the requirement on the strength of the effective optomechanical interaction, as quantified by the parameter $r=\frac{g^2 \tpulse}{\kappa}$, scales logarithmically with the initial occupation number $n_0$ of the mechanical oscillator. This tremendously eases the protocol's experimental realization, as neither $g$ nor $\tpulse$ can be arbitrarily increased---both for fundamental and technical reasons---, as we will show in \sref{sec:optimised}. Note that $n_0$ need not be equal to the mean bath occupation $\bar{n}$, but may be decreased by laser pre-cooling to improve the protocol's performance.
\subsubsection{Entanglement verification}
\label{sec-2-1-3}

To verify the successful creation of entanglement a red detuned laser pulse ($\detuning=\om$) is sent to the cavity where it resonantly drives the beam-splitter interaction, and hence generates a state swap between the mechanical and the optical mode. It is straightforward to show that choosing $\detuning=\om$ leads to a different set of Langevin equations which can be obtained from \eqref{eq:4} by dropping the Hermitian conjugation ($\dagger$) on the right-hand side. By defining modified mode functions $\alpha'_{\mathrm{in(out)}}=\alpha_{\mathrm{out(in)}}$ and corresponding light modes $A'_{\mathrm{in(out)}}$ one obtains input/output expressions in analogy to \eqref{eq:8}
\begin{subequations}
  \label{eq:13}
  \begin{align}
    \label{eq:13a}
    A'_{\mathrm{out}}&=-\e^{-G\tpulse} A'_{\mathrm{in}} +\rmi \sqrt{1-\e^{-2G\tpulse}}B_{\mathrm{in}},\\
    \label{eq:13b}
    B_{\mathrm{out}}&=\e^{-G\tpulse} B_{\mathrm{in}} - \rmi
    \sqrt{1-\e^{-2G\tpulse}}A'_{\mathrm{in}}.
  \end{align}
\end{subequations}
The pulsed state-swapping operation therefore also features an exponential scaling with $G\tpulse$. For $G\tpulse \rightarrow \infty$ the expressions above reduce to $A'_{\mathrm{out}}=-\rmi B_{\mathrm{in}}$ and $B_{\mathrm{out}}=\rmi A'_{\mathrm{in}}$, which shows that in this case the mechanical state---apart from a phase shift---is perfectly transferred to the optical mode. In the Schrödinger-picture this amounts to the transformation $|\varphi\rangle_{\mathrm{m}}|\psi\rangle_{\mathrm{l}}\rightarrow |\psi\rangle_{\mathrm{m}}|\varphi\rangle_{\mathrm{l}}$, where $\varphi$ and $\psi$ constitute the initial state of the mechanics and the light pulse, respectively.
The state-swap operation thus allows us to access mechanical quadratures by measuring quadratures of the light and therefore to reconstruct the state of the bipartite system via optical homodyne tomography. For this the protocol is operated in two steps: After the blue detuned pulse is reflected from the cavity, it is sent to a homodyne detection setup, where a quadrature $X_{\mathrm{l}}(\phi) =X_{\mathrm{l}}^{\mathrm{out}} \cos \phi + P_{\mathrm{l}}^{\mathrm{out}} \sin \phi$ ($\phi$ being the local oscillator phase) is measured. The same procedure is subsequently carried out for a red detuned pulse, measuring $X'_{\mathrm{l}}(\theta)$ (see \fref{fig:protocol}), which, for the case of a perfect state swap, yields the mechanical quadrature $X'_{\mathrm{l}}(\theta)=X_{\mathrm{m}}(\theta+\frac{\pi}{2})$. Here the rotation by $\frac{\pi}{2}$ is due to the phase shift from the swap operation. By repeating this process multiple times for different local-oscillator phases $(\phi_i,\theta_j)$, the quantum state of the bipartite optomechanical system can be reconstructed. Having obtained the full quantum state, entanglement can be analyzed by various means \cite{adesso_entanglement_2007}, \eg{}, by applying the EPR criterion from above.
\subsubsection{Optomechanical teleportation protocol}
\label{sec-2-1-4}
\label{sec:protocol}

As we have shown above, pulsed operation allows us to create EPR-type entanglement, which forms the central entanglement resource of many quantum information processing protocols \cite{braunstein_quantum_2005}. An immediate extension this scheme is an optomechanical continuous variables quantum teleportation protocol. The main idea of quantum state teleportation in this context is to transfer an arbitrary quantum state $|\psi_{\mathrm{in}}\rangle$ of a traveling wave light pulse onto the mechanical resonator, without any direct interaction between the two systems, but by making use of optomechanical entanglement. The scheme works in full analogy to the CV teleportation protocol for photons \cite{vaidman_teleportation_1994,braunstein_teleportation_1998}. Due to its pulsed nature it closely resembles the scheme used in atomic ensembles \cite{hammerer_teleportation_2005,sherson_quantum_2006} and it was recently also suggested in the context of levitated microspheres \cite{romero-isart_optically_2011} (see \cite{schwager_interfacing_2010} for an exhaustive description of a similar system comprising a nuclear-spin ensemble entangled with light):
A light pulse (A) is sent to the optomechanical cavity and is entangled with its mechanical mode (B) via the dynamics described above. Meanwhile a second pulse (V) is prepared in the state $|\psi_{\mathrm{in}}\rangle$, which is to be teleported. This pulse then interferes with A on a beam splitter. In the output ports of the beam splitter, two homodyne detectors measure two joint quadratures $P^{\mathrm{out}}_{\mathrm{l}}+X_{\mathrm{v}}$ and $X^{\mathrm{out}}_{\mathrm{l}}+P_{\mathrm{v}}$, yielding outcomes $m_{X}$ and $m_{P}$, respectively. This constitutes the analogue to the Bell measurement in the case of qubit teleportation and effectively projects previously unrelated systems A and V onto an EPR state \cite{bouwmeester_experimental_1997}. Note that both the second pulse and the local oscillator for the homodyne measurements must be mode-matched to A after the interaction; \ie{}, they must possess the identical carrier frequency as well as the same exponential envelope. The protocol is concluded by displacing the mirror in position and momentum by $m_{\mathrm{X}}$ and $m_{\mathrm{P}}$ according to the outcome of the Bell-measurement. This can be achieved by means of short light pulses, applying the methods described in \cite{vanner_pulsed_2011,cerrillo_pulsed_2011}. After the feedback the mirror is then described by \cite{braunstein_quantum_2005}
\begin{subequations}
  \label{eq:14}
  \begin{align}
    \label{eq:15}
    X_{\mathrm{m}}^{\mathrm{fin}}&=X_{\mathrm{m}}^{\mathrm{out}}+P^{\mathrm{out}}_{\mathrm{l}}+X_{\mathrm{v}},\nonumber\\
    &=X_{\mathrm{v}}+\left(\e^r-\sqrt{\e^{2r}-1}\right)(X_{\mathrm{m}}^{\mathrm{in}}-P^{\mathrm{in}}_{\mathrm{l}}),\\
    \label{eq:16}
    P_{\mathrm{m}}^{\mathrm{fin}}&=P_{\mathrm{m}}^{\mathrm{out}}+X^{\mathrm{out}}_{\mathrm{l}}+P_{\mathrm{v}},\nonumber\\
    &=P_{\mathrm{v}}+\left( \e^r-\sqrt{\e^{2r}-1} \right)(P_{\mathrm{m}}^{\mathrm{in}}-X^{\mathrm{in}}_{\mathrm{l}}),
  \end{align}
\end{subequations}
which shows that its final state corresponds to the input state plus quantum noise contributions. It is obvious from these expressions that the total noise added to both quadratures [second term in \eqref{eq:15} and \eqref{eq:16}, respectively] is equal to the EPR variance. Again, for large squeezing $r\gg 1$ the noise terms are exponentially suppressed and in the limit $r\rightarrow \infty$, where the resource state approaches the EPR state, we obtain perfect teleportation fidelity, \ie{}, $X^{\mathrm{fin}}_{\mathrm{m}}=X_{\mathrm{v}}$ and $P^{\mathrm{fin}}_{\mathrm{m}}=P_{\mathrm{v}}$. In particular this operator identity means, that \emph{all} moments of $X_{\mathrm{v}}$, $P_{\mathrm{v}}$ with respect to the input state $|\psi_{\mathrm{in}}\rangle$ will be transferred to the mechanical oscillator, and hence its final state will be identically given by $|\psi_{\mathrm{in}}\rangle$.

To verify the success of the teleportation one has to read out the mirror state after completing the feedback step. This can be achieved by applying tomography on the mechanical state as described in the previous section.
The overlap of the reconstructed state $\rho_{\mathrm{out}}$ and a pure input state $| \psi_{\mathrm{in}} \rangle$ then gives the teleportation fidelity $F = \langle \psi_{\mathrm{in}} | \rho_{\mathrm{out}} | \psi_{\mathrm{in}} \rangle$.
For coherent input states the fidelity is given by
\begin{equation}
  \label{eq:17}
  F=\left( 1+\frac{\depr}{2} \right)^{-1}.
\end{equation}
In order to beat the optimal classical strategy for transmission of quantum states (\ie{}, the optimal measure-and-prepare scheme), the achieved fidelity (averaged over all coherent states) must exceed \(F>1/2\) \cite{hammerer_quantum_2005}, which is equivalent to the condition for entanglement, $\depr<2$.
\subsection{Optimised protocol including imperfections}
\label{sec-2-2}
\label{sec:optimised}
\subsubsection{Perturbations}
\label{sec-2-2-1}
\label{sec:perturbations}

\label{sec-2-2-1-1}

In the previous section we found that in the ideal scenario the amount of entanglement essentially depends only on the coupling strength (or equivalently on the input laser power) and the duration of the laser pulse and that it shows an encouraging scaling, growing exponentially with $G\tpulse$. This in turn means that the minimal amount of squeezing needed to generate entanglement only grows logarithmically with the initial mechanical occupation $n_0$. In this section we will develop a more realistic scenario including thermal noise effects and full system dynamics, both of which will decrease the created entanglement. We will show, however, that under conditions already available in state-of-the-art optomechanical experiments, one can find an optimal working point such that the significance of these unwanted effects can sufficiently be suppressed.

\label{sec-2-2-1-2}

To extend the validity of the previous, simplified model, we now include the following additional dynamics: contributions from the beam-splitter Hamiltonian, higher order interactions beyond the adiabatic approximation, and decoherence effects due to mechanical coupling to a heat bath. In the following we will investigate their effect on our protocol and determine the parameter regime featuring maximal entanglement. The technical details of how we include them in our calculations will be shown in \sref{sec:completemodel}.

\label{sec-2-2-1-3}

Including the above-mentioned perturbations results in a final state which deviates from an EPR-entangled state. To minimise the extent of these deviations, the system parameters must obey the following conditions:
\begin{enumerate}
\item $\kappa\ll\om$ results in a sharply peaked cavity response and implies that the down-conversion dynamics is heavily enhanced with respect to the suppressed beam-splitter interaction.
\item $g \lesssim \kappa$ inhibits multiple interactions of a single photon with the mechanical mode before it leaves the cavity. This suppresses spurious correlations to the intracavity field. It also minimises pulse distortion and simplifies the protocol with regard to mode matching and detection.
\item $g \tpulse \gg 1$ is needed in order to create sufficiently strong entanglement. This is due to the fact that the squeezing parameter $r=(g/\kappa) g\tpulse$ should be large, while $g/\kappa$ needs to be small.
\item $\bar{n} \gamma \tpulse\ll 1$, where $\bar{n}$ is the thermal occupation of the mechanical bath, assures coherent dynamics over the full duration of the protocol, which is an essential requirement for observing quantum effects. As the thermal occupation of the mechanical bath may be considerably large even at cryogenic temperatures, this poses (for fixed $\gamma$ and $\bar{n}$) a very strict upper limit to the pulse duration $\tpulse$.
\end{enumerate}

Note however that not all of these inequalities have to be fulfilled equally strictly, but there rather exists an optimum which arises from balancing all contributions. It turns out that fulfilling (4) is critical for successful teleportation, whereas (1)--(3) only need to be weakly satisfied. Taking the above considerations into account, we find a sequence of parameter inequalities
\begin{equation}
  \label{eq:18}
  \bar{n}\gamma \ll \frac{1}{\tpulse}\ll g \ll \kappa \ll \om,
\end{equation}
which defines the optimal parameter regime. In \sref{sec:motivation} we assumed the first two conditions to be well satisfied and we neglected the existence of mechanical decoherence. If we now take into account that the mechanical oscillator couples to a heat bath with an effective decoherence rate $\bar{n} \gamma$, we find that increasing the pulse duration to values larger than the mechanical coherence time will drastically decrease entanglement. This results in an upper bound for entanglement, as now both the interaction strength and the pulse duration, and therefore also the squeezing parameter $r=(g/\kappa)g\tpulse$, are bounded from above.

Dividing \eqref{eq:18} by $\gamma$ and taking a look at the outermost condition $\bar{n} \ll \qm$, where $\qm=\om/\gamma$ is the mechanical quality factor, we see that the ratio $\qm/\bar{n}$ defines the range which all the other parameters have to fit into. It is intuitively clear, that a high quality factor and a low bath occupation number, and consequently a low effective mechanical decoherence rate, are favourable for the success of the protocol. Equivalently, we can rewrite the occupation number as $\bar{n}=k_{\mathrm{B}}T_{\mathrm{bath}}/\hbar \om$ and therefore find $k_{\mathrm{B}}T_{\mathrm{bath}}/\hbar \ll \qm\cdot\om$, where now the $Q\cdot f$-product ($f=\om/2\pi$) has to be compared to the thermal frequency of the bath.
Let us consider a numerical example: For a temperature $T_{\mathrm{bath}}\approx 100\,\mathrm{mK}$ the left-hand side gives $k_{\mathrm{B}}T_{\mathrm{bath}}/\hbar\approx 2\pi\cdot 10^{9}\,\mathrm{Hz}$. The $Q \cdot f$-product consequently has to be several orders of magnitude larger to successfully create entanglement. As current optomechanical systems feature a $Q\cdot f$-product of $2\pi \cdot 10^{11}\,\mathrm{Hz}$ and above \cite{cole_phonon-tunnelling_2011,ding_high_2010,eichenfield_optomechanical_2009,safavi-naeini_electromagnetically_2011}, this requirement seems feasible to meet.
Note that in an experiment $T_{\mathrm{bath}}$ will often depend on the input laser power, as scattered light can heat up the cryogenic environment. Hence,  for a given bath occupation the coupling strength may be limited for technical reasons.

\label{sec-2-2-1-4}

In order to find the optimal working point, it is convenient to introduce the following dimensionless parameters: the sideband-resolution parameter $\eta$, the adiabaticity parameter $\xi$, and the ratio of pulse length to mechanical coherence time $\epsilon$. They are given by
\begin{align*}
  \eta=\kappa/\om,&&\xi=g/\kappa,&&\epsilon=\gamma \tpulse.
\end{align*}
From \eqref{eq:18} it follows that $\eta \ll 1$, $\xi\ll 1$ and $\epsilon\ll 1/\bar{n}$. Each of those small parameters can be used to realise a perturbative expansion of the additional dynamics listed above. The perturbative solutions can then be used to calculate the EPR variance and optimise the resulting expressions.
\subsubsection{Optimization}
\label{sec-2-2-2}
\label{optimization}

\label{sec-2-2-2-1}

\begin{figure}
  \begin{minipage}{\columnwidth}\rlap{(a)}{\includegraphics[width=.95\columnwidth]{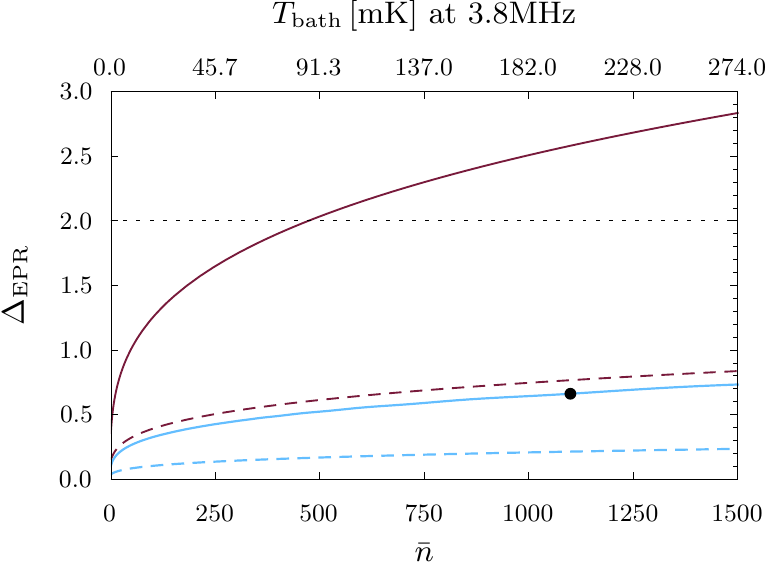}}\end{minipage}\\
  \vspace{1em}
  \begin{minipage}{.5\columnwidth}\rlap{(b)}{\includegraphics[width=.95\columnwidth]{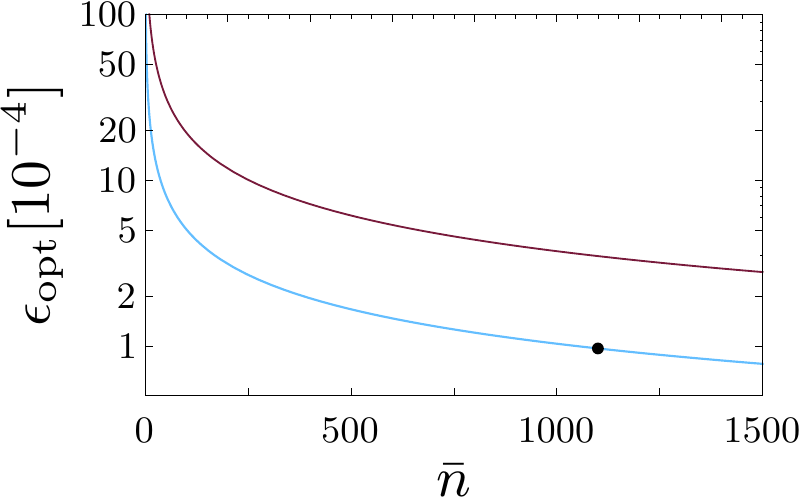}}\end{minipage}%
  \hfill%
  \begin{minipage}{.5\columnwidth}\rlap{(c)}{\includegraphics[width=.95\columnwidth]{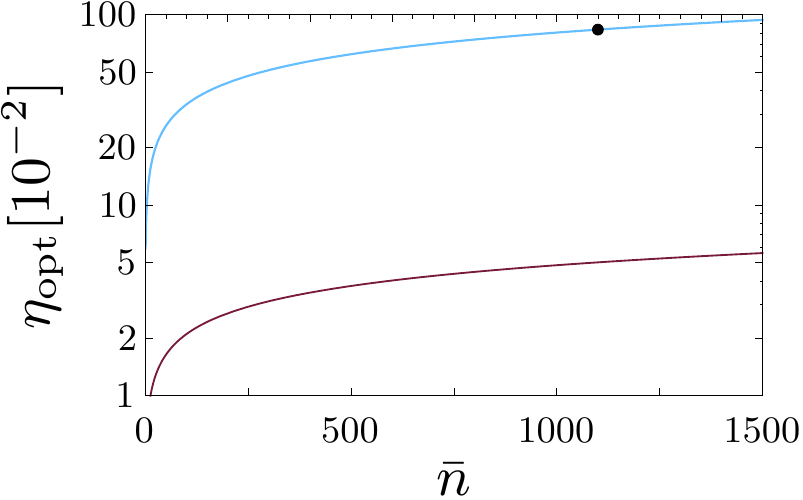}}\end{minipage}\\
  \vspace{1em}
  \begin{minipage}{.5\columnwidth}\rlap{(d)}{\includegraphics[width=.95\columnwidth]{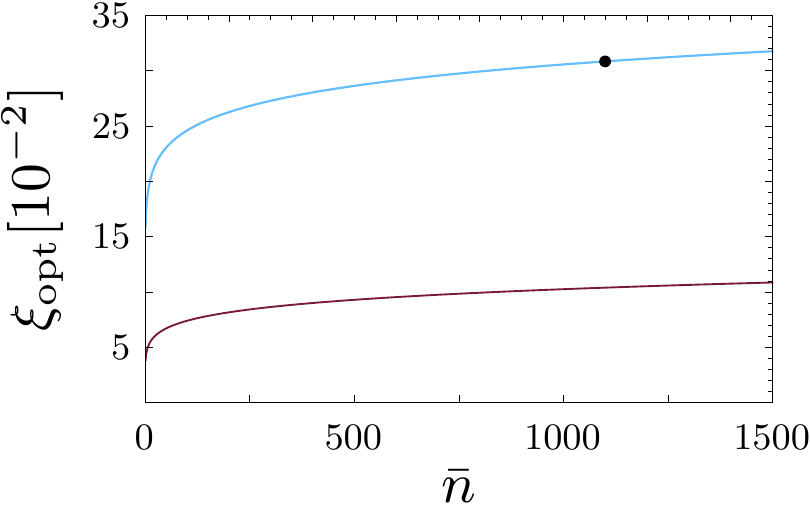}}\end{minipage}%
  \hfill%
  \begin{minipage}{.5\columnwidth}\rlap{(e)}{\includegraphics[width=.9\columnwidth]{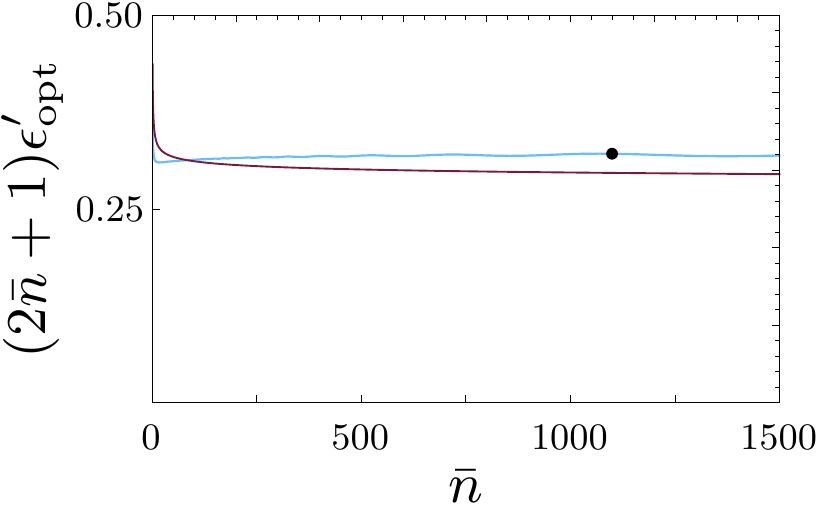}}\end{minipage}%
  \caption{(Color online) Optimised parameters for $\qm=10^7$ and $n_0=50$ [red (dark) line], and $\qm=10^5$ and $n_0=0$ [blue (light) line], where $n_0$ is the initial mechanical occupation and $\bar{n}$ is the mean bath occupation. This corresponds to the two cases of large $\qm$ with moderate pre-laser-cooling, and lower $\qm$ with pre-cooling into the ground state. Clearly, entanglement creation is possible in both cases. (a) Minimal $\depr$ as a function of $\bar{n}$. The upper axis gives the corresponding bath temperature for a oscillator with a resonance frequency of 3.8\,MHz. To each point corresponds a triple $\epsilon_{\mathrm{opt}},\eta_{\mathrm{opt}},\xi_{\mathrm{opt}}$ [(b)--(d)] for which the minimal value is realised. The dotted, black line shows the upper bound up to which entanglement is present. The black dots marks the values for $\bar{n}=1100$ ($T=200\,\mathrm{mK}$), which coincide with the values given in the first row of table~\ref{tab:specval}. The dashed lines show the respective thermal noise contribution $(2 \bar{n}+1)\epsilon_{\mathrm{opt}}$ to $\depr$ [see \eqref{eq:19}]. (b)--(d) Values of $\epsilon,\eta,\xi$ which optimise $\depr$ for a given $\bar{n}$. (e) Shows the relative amount of noise induced by the coupling to the mechanical environment ($\epsilon'_{\mathrm{opt}}=\epsilon_{\mathrm{opt}}/\depr$).}
  \label{fig:optimization}
\end{figure}
As illustrated above, we expect---for fixed values of $\bar{n}$ and $\qm$---to find optimal values for the remaining parameters $\epsilon,\xi$ and $\eta$. The maximal possible entanglement will ultimately be set by $\bar{n}$ and $\qm$, which will also constitute hard boundaries in typical experiments.

Figure \ref{fig:optimization} shows results of this optimization for different $Q$-factors, plotted against the thermal occupation number of the mechanical bath. Figure (a) shows the minimal value of $\depr$ for a given $\qm$, and (b)--(d) show the corresponding optimal values for $\eta_\mathrm{opt}$, $\xi_\mathrm{opt}$ and $\epsilon_\mathrm{opt}$. As expected the noise contribution from the mechanical bath is found to be the most critical. As we show in \ref{sec:fullequations} the EPR variance including thermal noise can be expressed by
\begin{multline}
  \label{eq:19}
  \depr =\left[\Delta(X^{\mathrm{out}}_{\mathrm{m}}+P^{\mathrm{out}}_{\mathrm{l}})\right]^2\\
  +\left[\Delta(P^{\mathrm{out}}_{\mathrm{m}}+X^{\mathrm{out}}_{\mathrm{l}})\right]^2 + (2\bar{n}+1)\epsilon,
\end{multline}
where $X^{\mathrm{out}}_i$ and $P^{\mathrm{out}}_i$ here denote the solutions for $\gamma=0$.
The dashed curves in \fref{fig:optimization}~(a) illustrate the noise contribution $(2\bar{n}+1)\epsilon_\mathrm{opt}$ of the thermal bath. As shown above, this quantity is added to the unperturbed EPR variance, and thus the system can only exhibit strong entanglement if its value is far below two.
Note that working at the optimal point keeps the fraction of thermal noise in $\depr$ approximately constant over a wide range of $\bar{n}$. This is shown in \fref{fig:optimization}~(e), where we defined $\epsilon'_{\mathrm{opt}}=\epsilon_{\mathrm{opt}}/\depr$.

As we have seen in the previous section, the EPR variance depends on the occupation number of the oscillator at the initial time $t=0$. Due to this, the entanglement can be drastically increased by pre-cooling the mechanics by means of laser cooling before starting the actual protocol. \Fref{fig:optimization} shows that it is thus possible to create an entangled state even for a fairly large bath occupation. This works due to short pulse durations, during which the mechanical decoherence is small.

Taking a look at figure (c) we note that the sideband-resolution shows rather large optimal values near unity, especially for increasing occupation numbers. This indicates that the beam-splitter dynamics only weakly disturbs the entangling interaction.

Table~\ref{tab:specval} gives a list of the optimized key experimental parameters for two existing optomechanical structures. The values in the first row correspond to the black dots shown in \fref{fig:optimization}.

Over and above the fundamental imperfections, technical losses, such as mode-mismatch and detector inefficiencies, additionally decrease entanglement. They can all collectively be described as (passive) beam-splitter losses adding vacuum noise to the optical signal. They will however, never completely break entanglement, as long as the overall loss is smaller than unity. Noise contributions of this type can easily be accounted for by adding appropriate noise terms to \eqref{eq:8}.

Finally let us compare the amount of entanglement created in the pulsed and continuous-wave schemes \cite{genes_robust_2008} in terms of the logarithmic negativity $E_{\mathcal{N}}$ \cite{adesso_entanglement_2007} for the parameters used in the first row of table~\ref{tab:specval} ($\qm=10^5$, $\bar{n}=1100$, $n_0=0$). In the case of continuous driving one finds the maximal negativity close to the instability region and for a detuning of around $\detuning \approx \om$, yielding $E_{\mathcal{N}}\approx 0.4$. For the pulsed protocol the optimisation yields a much larger value of $E_{\mathcal{N}}\approx 1.2$. Note that $E_{\mathcal{N}}$ is a logarithmic quantity.

\begin{table*}
  \caption{\label{tab:specval}Specific \emph{optimal} values for an $\mathrm{Al}_{x}\mathrm{Ga}_{1-x}\mathrm{As}$ structure (3.8\,MHz) \cite{cole_megahertz_2010} and an Si optomechanical crystal (3.7\,GHz) \cite{chan_laser_2011}. The whole set is fully determined by $\epsilon_{\mathrm{opt}}$, $\eta_{\mathrm{opt}}$, $\xi_{\mathrm{opt}}$. The value for the mean power $P=\hbar \omega_\mathrm{l} N_\mathrm{ph}/\tpulse$ is obtained using (\ref{eq:2}).
    (The $x$ in the formula above is a number between 0 and 1 to indicate a ternary alloy between GaAs and AlAs.)}
  \begin{ruledtabular}\begin{tabular}{rrrrrdddrdr}
      $\om/2\pi$ & $\qm$ & $T_{\mathrm{bath}}$ & $\bar{n}$ & $n_{0}$ & g_{0}/2\pi & \kappa_\mathrm{opt}/2\pi & \tpulse_\mathrm{opt} & $P_\mathrm{opt}$ & g_\mathrm{opt}/2\pi & $\depr$ \\
      \hline
      3.8\,MHz & $10^5$ & 200\,mK & 1100 & 0.0 & 4.8\,\mathrm{Hz} & 3.2\,\mathrm{MHz} & 2.5\,\mathrm{\mu s} & $30\,\mathrm{mW}$ & 0.97\,\mathrm{MHz} & 0.7 \\
      3.7\,GHz & $10^5$ & 200\,mK & 0.7 & 0.7 & 910.0\,\mathrm{kHz} & 0.26\,\mathrm{GHz} & 0.41\,\mathrm{\mu s} & $6\,\mathrm{\mu W}$ & 0.032\,\mathrm{GHz} & 0.1 \\
      3.7\,GHz & $10^5$ & 1\,K & 3.7 & 3.7 & 910.0\,\mathrm{kHz} & 0.31\,\mathrm{GHz} & 0.30\,\mathrm{\mu s} & $8\,\mathrm{\mu W}$ & 0.040\,\mathrm{GHz} & 0.5
    \end{tabular}\end{ruledtabular}
\end{table*}
\section{Detailed model}
\label{sec-3}
\label{sec:completemodel}
\subsection{Linearizing the dynamics}
\label{sec-3-1}
\label{sec:linearization}

As was shown in \cite{law_interaction_1995}, the radiation pressure interaction is inherently non-linear. In current micromechanical systems however, the single-photon coupling $g_0$ is very weak (the best values up to date are on the order of $g_0/\kappa \approx 0.001$ \cite{safavi-naeini_electromagnetically_2011}), and has to be enhanced by means of a strong optical pump field. It is well known that in the case of a strong continuous-wave light field, the steady-state dynamics of the system is approximately linear. We will show in the following, that this also holds in a (long) pulsed scheme. The linearization process follows the same general idea as in the steady-state regime, it is, however, slightly more involved due to the explicit time dependence of the Hamiltonian.

We consider a laser pulse with a fixed number of photons $N_\mathrm{ph}$ and an envelope function $\varepsilon(t)$, which is normalised in the sense that $\intdb{0}{\tpulse}{t}|\varepsilon(t)|^2=1$. Its head and tail are assumed to be smooth and its amplitude should be constant $\varepsilon(t)\approx 1/\sqrt{\tpulse}$ for the most part of $\tpulse$.
The full Hamiltonian for the system, including the laser driving term and the non-linear radiation pressure interaction \cite{law_interaction_1995} is then given by
\begin{multline}
  \label{eq:20}
  H(t)= \om^{} \amd\am + \Delta_0^{}\acd\ac + \\
  g_0^{}\acd\ac \left( \am + \amd \right) +\rmi E(t)\left( \ac - \acd \right),
\end{multline}
where $\Delta_0=\omega_\mathrm{c,0}-\omega_{\mathrm{l}}$ is the detuning for the case of a cavity with fixed length and $E(t)=\sqrt{2\kappa N_{\mathrm{ph}}} \varepsilon(t)$ is the driving strength. In the Appendix we show that we can eliminate the driving term by going into a (time-dependent) displaced picture. The transformed Hamiltonian then takes the form
\begin{multline}
  \label{eq:21}
  \bar{H}(t)=\om^{} \amd \am+\detuning^{}(t)\acd \ac +g_0^{}\acd \ac(\am+\amd) \\
  +g_0^{}\left(\alpha^*(t)\ac+\alpha(t) \acd\right)(\am+\amd),
\end{multline}
where the effective detuning $\detuning$ and the mean cavity field $\alpha$ now depend explicitly on time. The resulting expressions (see the Appendix) are essentially the same as found in the steady-state case (see for example \cite{genes_ground-state_2008}).
The non-linear term in \eqref{eq:21} can be neglected whenever $\abs{\alpha}\gg 1$, which is true for sufficiently strong driving $\abs{E}$ and will be the case for the greatest part of the pulse duration.
By assuming $\alpha(t)$ to be real and by introducing the effective optomechanical coupling constant $g(t)=g_0\abs{\alpha(t)}$, the procedure leaves us with a linear Hamiltonian in the form of \eqref{eq:1}.

Note that for the case of $\detuning\approx \om$ the relative frequency shift induced by radiation pressure will be of the order of $O (g/\om)^2$, and therefore small.
Consequently [together with the assumption that \(\varepsilon(t)\approx 1/\sqrt{\tpulse}\)] we will in the following drop the explicit time dependence of the effective detuning and the effective coupling strength.
\subsection{Solving the full system}
\label{sec-3-2}
\label{sec:fullequations}

The full Langevin equations \eqref{eq:3} resulting from the linearised Hamiltonian (including the beam-splitter interaction and mechanical decoherence) can be rewritten in the compact form
\begin{align}
  \label{eq:19-1}
  \frac{\rmd}{\rmd t}\Rop(t) = (S-D)\Rop(t) - \sqrt{2D}\Rin(t),
\end{align}
where $\Rop=(\am,\amd,\ac,\acd)$ and correspondingly $\Rin$ denotes the input noise.
$S$ and $D$ are matrices comprising the respective coefficients. To solve this set of equations we apply the Laplace transformation, introducing $\bar{\Rop}(s) = \mathcal{L}[\Rop](s)$, with $\mathcal{L}[f](s) = \intdb{0}{\infty}{t} \e^{-s t} f(t)$. Solving for $\Rop$ we obtain
\begin{align}
  \label{eq:22}
  \bar{\Rop}(s) = \bar{M}(s)\left(\Rop(0) - \sqrt{2D}\bar{\Rop}_{\mathrm{in}}(s)\right),
\end{align}
and thus
\begin{align}
  \label{eq:23}
  \Rop(\tpulse) = M(\tpulse)\Rop(0) - \left( M*\sqrt{2D}\Rin \right)(\tpulse),
\end{align}
where $\bar{M}(s) = (s \mathbf{1} - S+D)^{-1}$, $M(\tpulse)=\mathcal{L}^{-1}[\bar{M}](\tpulse)$ and $*$ denotes the convolution integral $(f*g)(t)=\intdb{0}{t}{s}g(s)f(t-s)$.
In the case of a bipartite system it is possible to find an exact expression for $M$ for arbitrary parameters. The obtained solution, however, is very tedious and will not be presented here. We proceed as follows: We separate the mechanical decoherence in a perturbative approach (\ie{}, we expand $M$ in powers of $\epsilon$), while the other dynamics will be treated exactly. This allows us to find input/output relations corresponding to \eqref{eq:8} and to calculate the EPR variance for the full system.

We established in \sref{optimization} that for the protocol to work we require the effective mechanical decoherence time to be much larger than the duration of the light pulse, \ie{}, $\epsilon \,\bar{n}\ll 1$. We emphasise that $\epsilon$ is the smallest of all parameters, and the coherent evolution will only be negligibly perturbed by the coupling to the mechanical bath. We will therefore only keep terms $O(\epsilon \bar{n})$ while neglecting $O(\epsilon)$. Based on this premise we simplify \eqref{eq:23} twofold: Firstly we drop the mechanical damping from the first term, as it gives corrections on the order of $O(\epsilon)$ only; thus $M\approx M|_{\gamma=0}$. This amounts to dropping the term $-\frac{\gamma}{2}\, p_{\mathrm{m}}$ in \eqref{eq:3b}. Secondly, we approximate the mechanical noise contribution (second term) by only keeping the free, harmonic evolution, while neglecting their coupling to the optical mode. This coupling is due to a second-order process, and is therefore suppressed by an additional factor of $\xi$. Note that by doing so we overestimate the effect of mechanical noise, as it contributes to the creation of optomechanical correlations when subject to the coherent dynamics. The complete noise term entering in the evolution of the mechanical variables then takes the form
\begin{equation}
  \label{eq:32}
  \rmi \sqrt{\gamma}\intdb{0}{\tpulse}{s} \e^{-\rmi \om s} f(\tpulse-s) \rdef \sqrt{\frac{\gamma \tpulse}{2}}\left( F_{\mathrm{s}} + \rmi F_{\mathrm{c}} \right),
\end{equation}
where we introduced (co)sine components $F_\mathrm{(c)s}$ of the Brownian force. We can therefore write $\am (\tpulse) \approx \am(\tpulse)\big|_{\gamma=0} + \sqrt{\epsilon/2}\, (F_{\mathrm{s}}+\rmi F_{\mathrm{c}})$,
and consequently
\begin{subequations}
  \label{eq:33}
  \begin{align}
    X_{\mathrm{m}}^{\mathrm{out}} &\approx X_{\mathrm{m}}^{\mathrm{out}}\Big|_{\gamma=0} + \sqrt{\epsilon}\, F_{\mathrm{s}},\\
    P_{\mathrm{m}}^{\mathrm{out}} &\approx P_{\mathrm{m}}^{\mathrm{out}}\Big|_{\gamma=0} + \sqrt{\epsilon}\, F_{\mathrm{c}},
  \end{align}
\end{subequations}
while we neglect the mechanical noise contribution to the optical mode, \ie{}, $A_{\mathrm{out}}\approx A_{\mathrm{out}}|_{\gamma=0}$.
Note that from the commutation relation of the Brownian noise term, $[f(t+s),f(t)]=\frac{\rmi}{\om}\delta'(s)$ \cite{gardiner_quantum_2004}, it follows that $[F_{\mathrm{s}},F_{\mathrm{c}}]=\rmi + O(1/\om \tpulse)$ and $[F_{i},F_{i}]= O(1/\om \tpulse)$. The perturbed variables (\ref{eq:33}) therefore approximately obey canonical commutation relations $[X_{\mathrm{m}}^{\mathrm{out}},P_{\mathrm{m}}^{\mathrm{out}}]\approx \rmi(1+\epsilon)$. Using (\ref{eq:33}) together with the correlation functions $\langle F_{i}F_{i} \rangle =  \bar{n} + \frac{1}{2}$ leads to \eqref{eq:19}.
As we have separated the mechanical noise terms from the other dynamics, we will always assume that $M\approx M|_{\gamma=0}$ and drop the $\gamma$ dependence for the rest of this section.

We now use \eqref{eq:23} together with the definitions of $A_{\mathrm{out}}$ \eqref{eq:7b} and $B_{\mathrm{out}}$ to obtain input/output equations similar to \eqref{eq:8}, but for the full system dynamics. The resulting expressions are of the form
\begin{multline*}
  B_{\mathrm{out}}=c_1B_{\mathrm{in}}+c_2B_{\mathrm{in}}^{\dagger}+c_3\ac(0)+c_4\acd(0)\\
  +c_5\intdb{0}{\tpulse}{s}\alpha_{\mathrm{in},1}(s)\ain(s)+c_6\intdb{0}{\tpulse}{s}\alpha_{\mathrm{in},2}^{*}(s)\aind(s),
\end{multline*}
with a similar expression for $A_{\mathrm{out}}$. The coefficients $c_i$ as well as the light modes $\alpha_{\mathrm{in},i}$ are determined by the system dynamics [given by \(M(t)\)] and the light mode $\alpha_{\mathrm{out}}$ selected from the output field. Note that these expressions are valid for $\gamma=0$ only and thus have to be used in conjunction with \eqref{eq:33} to account for mechanical noise.

Following the treatment in \sref{sec:entanglement} one easily finds the corresponding EPR variance, which now includes noise terms from the initial intracavity field and the extra light modes (both assumed to be in vacuum). The latter contributions are given by the overlap of the different light modes $\intdb{0}{\tpulse}{t}\alpha_{\mathrm{in},i}(t) \,\alpha_{\mathrm{in},j}^{*}(t)$.

The resulting expression is an involved function of $\epsilon,\eta$ and $\xi$ and is not presented here. Numerical minimization with respect to those three variables for fixed values of $\bar{n}$ and $\qm$ yields the results presented in \fref{fig:optimization}.
\section{Conclusions}
\label{sec-4}

We have developed a scheme to create and---due to its pump--probe operation---unambiguously verify EPR entanglement in optomechanical systems. Additionally its application as an entanglement resource in quantum teleportation was discussed. Finally, by optimizing the experimental parameters we showed that the suggested protocol is feasible with state-of-the-art optomechanical devices.

\begin{acknowledgments}
  We thank the European Commission (MINOS, QESSENCE), the European Research Council (ERC StG QOM), the Austrian Science Fund (FWF) (START, SFB FoQuS), and the Centre for Quantum
  Engineering and Space-Time Research (QUEST) for support. W.\ W.\ acknowledges support from the Alexander von Humboldt Foundation. S.\,G.\ H.\ is a member of the FWF Doctoral Programme CoQuS (W1210). We thank G.\,D.\ Cole, N.\ Kiesel, M.\,R.\ Vanner  and A.\ Xuereb for useful discussion.
\end{acknowledgments}

\appendix
\section{Transformed Hamiltonian}
\label{sec-5}
\label{app:hamiltonian}

Starting from the Hamiltonian \eqref{eq:20} we write down the standard quantum-optical master equation for a damped cavity mode
\begin{align}
  \label{eq:24}
  \dot{\rho}=-\rmi \left[ H,\rho \right] + \kappa \left( 2 \ac\rho \acd - \ac \acd \rho -\rho \ac \acd  \right),
\end{align}
while we neglected the mechanical decoherence terms as we are only interested in times far within the coherence time of the oscillator.
In order to eliminate the driving field $E(t)$ we go into a displaced picture
\begin{equation}
  \label{eq:25}
  \bar{\rho}=D_{\mathrm{c}}(\alpha)D_{\mathrm{m}}(\beta)\rho D^{\dagger}_{\mathrm{c}}(\alpha)D^{\dagger}_{\mathrm{m}}(\beta),
\end{equation}
with displacement operators $D_i(\alpha)=\exp(\alpha a_i^{\dagger} + \alpha^{*} a_{i})$. The time-dependent, complex amplitudes $\alpha=\alpha(t)$ and $\beta=\beta(t)$ give the mean displacements due to the laser drive and will be determined in the following.

The transformed master equation can again be written in the form of \eqref{eq:24} by substituting $\rho\rightarrow \bar{\rho}$ and $H\rightarrow \bar{H}$. The Hamiltonian $\bar{H}$ is then given by
\begin{multline}
  \label{eq:26}
  \bar{H}
  =\om \amd \am+[\Delta_0+g_0(\beta+\beta^*)]a_c^\dagger a_c\\
  +g_0(\alpha^*a_c+\alpha a_c^\dagger)(\am+\amd)+g_0a_c^\dagger a_c(\am+\amd)\\
  +\Big\{\big[\rmi\dot\alpha+(\rmi\kappa+\Delta_{0})\alpha+g_0(\beta+\beta^*)\alpha-\rmi E\big]\acd + \mathrm{h.\,c.}\Big\}\\
  +\bigg\{\left[ \rmi\dot\beta + \om \beta+g_0|\alpha|^2 \right]\amd +\mathrm{h.\,c.}\bigg\}.
\end{multline}
The first two lines constitute the new Hamiltonian of the system, and the last two lines describe the mean (classical) cavity and mirror amplitude, respectively. We can make these terms disappear by choosing $\alpha$ and $\beta$ such that they fulfill the following set of coupled, non-linear differential equations
\begin{subequations}
  \label{eq:27}
  \begin{align}
    \label{eq:27a}
    \dot\alpha&=\left\{\rmi\left[\Delta_0+g_0(\beta+\beta^*)\right]-\kappa\right\}\alpha+E,\\
    \label{eq:27b}
    \dot\beta&= \rmi\om\beta+\rmi g_0|\alpha|^2.
  \end{align}
\end{subequations}
We seek solutions to these equations for initial conditions $\alpha(0)=\beta(0)=0$ in terms of the driving field $E(t)$. Due to their non-linear nature, no exact closed-form solution will exist in general and we will therefore look for approximate solutions under the assumptions we made in \sref{sec:optimised}. We formally integrate equation \eqref{eq:27b} to find
\begin{equation}
  \label{eq:28}
  \beta(t) = \rmi g_0 \intdb{0}{t}{s} \e^{\rmi \om s} \abs{\alpha(t-s)}^2.
\end{equation}
Under the assumption that $\kappa\ll\om$ and given that $E(t)$ varies sufficiently slowly, we also expect $\abs{\alpha(t)}^2$ to be a slowly varying function on the time scale of $1/\om$. We will check this for consistency at the end of this section. For this case we can use the adiabatic solution
\begin{equation}
  \label{eq:29}
  \beta(t)\approx - \frac{g_0}{\om} \abs{\alpha(t)}^2.
\end{equation}
Plugging this into \eqref{eq:27a} and introducing the effective detuning
\begin{equation}
  \label{eq:30}
  \begin{split}
    \detuning(t)&=\Delta_0-g_0(\beta(t)+\beta^{*}(t))\\
    &=\Delta_0-\frac{2g_0^2}{\om}\abs{\alpha(t)}^2
  \end{split}
\end{equation}
we find the solution
\begin{equation}
  \label{eq:31}
  \alpha(t)\approx \frac{1-\e^{-(\rmi \detuning+\kappa)t}}{\rmi\detuning+\kappa}E(t) \left( 1+\delta \right)\approx \frac{E(t)}{\rmi\detuning+\kappa},
\end{equation}
where $\delta$ is a correction, which is small if $E(t)$ varies slowly on a timescale of $1/\kappa$. More precisely, one can show that a crude upper bound is given by $|\delta(t)| <\sup_{s\in (0,t)}\frac{1}{\kappa} \frac{|\dot{E}(s)|}{\abs{E(t)}}$,
which must be much smaller than unity. Also, as we assume that $\kappa \tpulse \gg 1$, we neglect the term $\e^{-\kappa t}$, as this only contributes at the very beginning of the pulse. The approximations made in \eqref{eq:31} amount to assuming that the slope of the pulse is small enough (with respect to $\kappa$) that it does not experience distortion due to the finite cavity linewidth. Throughout this derivation we have assumed that $\detuning(t)\approx \detuning$ is approximately constant in time, which is well fulfilled for the parameter regime $O(g/\om)^2 \ll 1$ that we are concerned with. This can be made exact if we assume the laser detuning to be locked with respect to the effective cavity resonance frequency.
Having obtained a solution for $\alpha(t)$, we can go back to test the self-consistency of our derivation of \eqref{eq:29}, where we required $\frac{\rmd |\alpha|^2}{\rmd t}\ll \om$. In the case $g\ll\kappa\ll\om$ we find that $\frac{\rmd |\alpha|^2}{\rmd t}\approx\frac{1}{\om^2}\frac{\rmd |E|^2}{\rmd t}\ll \om$, giving an additional condition on the pulse shape.
Taking all these considerations into account we arrive at the linearised Hamiltonian \eqref{eq:21}.

\bibliography{TeleportationPaper}

\end{document}